# RZA-NLMF algorithm based adaptive sparse sensing for realizing compressive sensing problems

Guan Gui[#], Li Xu[*], and Fumiyuki Adachi[#]

[#]Department of Communications Engineering, Graduate School of Engineering, Tohoku University, Sendai, 980-8579 Japan

[*]Faculty of Systems Science and Technology, Akita Prefectural University, Akita, 015-0055 Japan

**Abstract**

Nonlinear sparse sensing (NSS) techniques have been adopted for realizing compressive sensing in many applications such as Radar imaging. Unlike the NSS, in this paper, we propose an adaptive sparse sensing (ASS) approach using reweighted zero-attracting normalized least mean fourth (RZA-NLMF) algorithm which depends on several given parameters, i.e., reweighted factor, regularization parameter and initial step-size. First, based on the independent assumption, Cramer Rao lower bound (CRLB) is derived as for the trademark of performance comparisons. In addition, reweighted factor selection method is proposed for achieving robust estimation performance. Finally, to verify the algorithm, Monte Carlo based computer simulations are given to show that the ASS achieves much better mean square error (MSE) performance than the NSS.

## 1. Introduction

Compressive sensing [1], [2] has been attracting high attentions in compressive Radar/sonar sensing [3], [4] due to many applications such as civilian, military, and biomedical. The main task of CS problems can be divided into three aspects as follows: 1) sparse signal learning: The basic model suggests that natural signals can be compactly expressed, or efficiently approximated, as a linear combination of prespecified atom signals, where the linear coefficients are sparse (i.e., most of them zero); 2) random measurement matrix design. It is important to make a sensing matrix which allows recovery of as many entries of unknown signal as possible by using as few measurements as possible

Sensing matrix should satisfy the conditions of incoherence and restricted isometry property (RIP) [5]. Fortunately, some special matrices (e.g., Gaussian matrix and Fourier matrix) have been reported that they are satisfying RIP in high probably; 3) sparse reconstruction algorithms. Based on previous two steps, many sparse reconstruction algorithms have been proposed to find the suboptimal sparse solution.

It was well known that the CS provides a robust framework that can reduce the number of measurements required to estimate a sparse signal. Many NSS algorithms and their variants have been proposed to deal with CS problems. They mainly fall into two basic categories: convex relaxation (basis pursuit de-noise, BPDN [6]) and greedy pursuit (orthogonal matching pursuit, OMP [7]). Above NSS based CS methods are either high complexity or low performance, especially in the case of low signal-to-noise (SNR) regime.

In this paper, we propose an adaptive sparse sensing (ASS) method using reweighted zero-attracting normalized mean fourth error algorithm (RZA-NLMF) [8] to solve the CS problems. Different from NSS methods, each observation and corresponding sensing signal vector will be implemented by the RZA-NLMF algorithm to reconstruct the sparse signal during the process of adaptive filtering. The effectiveness of our proposed method is confirmed via computer simulation when comparing with NSS.

The remainder of the paper is organized as follows. Basic CS problem is introduced and typical NSS method is presented in Section 2. In section 3, ASS using RZA-NLMF algorithm is proposed for solving CS problems and its derivation process is highlighted. Computer simulations are given in Section 4 in order to evaluate and compare performances of the proposed ASS method. Finally, our contributions are summarized in Section 5.

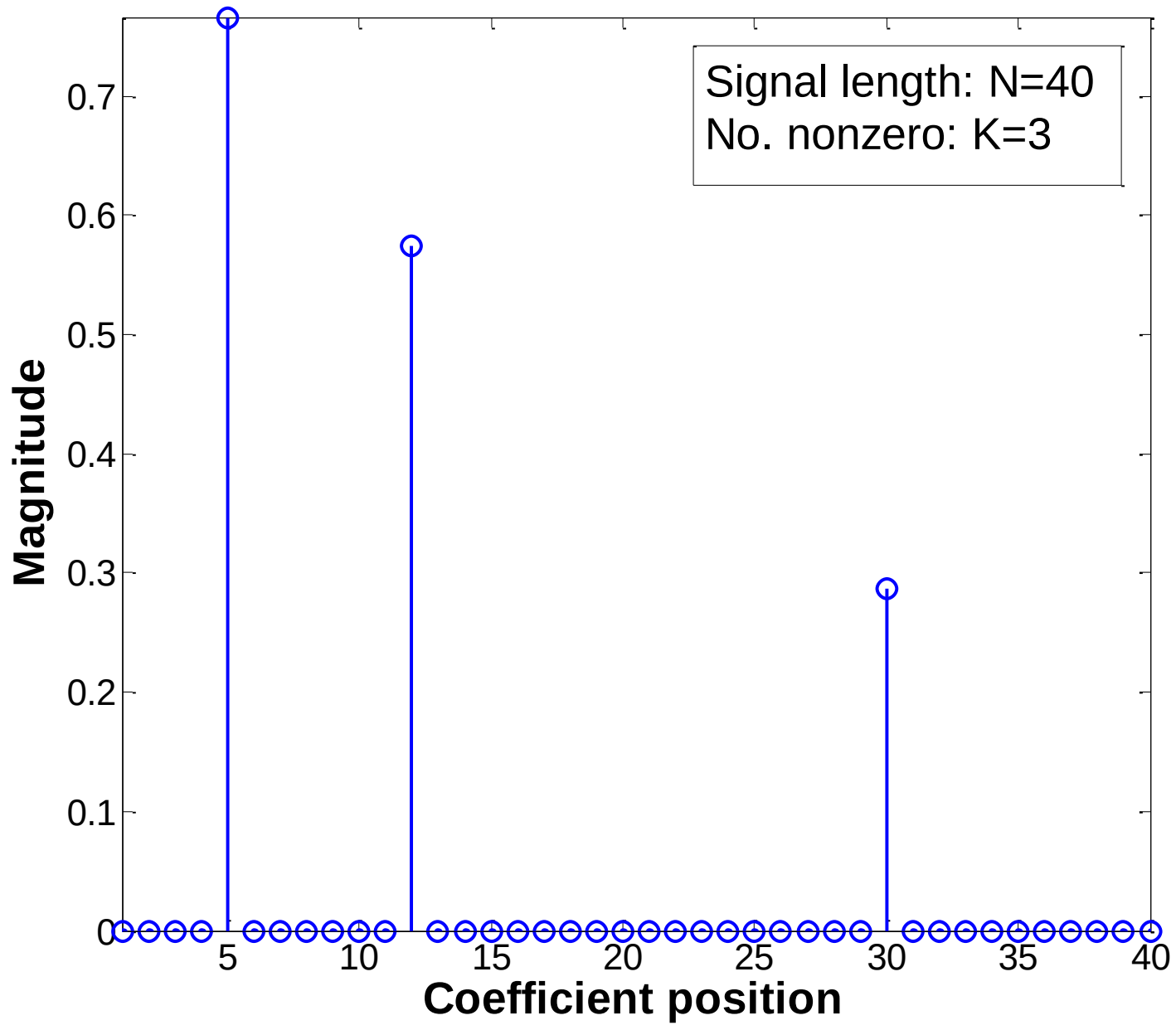

**Figure 1. A typical example of sparse structure signal.**

## 2. Nonliner sparse sensing

Assume a finite-length discrete signal vector $\boldsymbol{s}=[s_1,s_2,\cdots,s_N]^T$ can be sparse represented in a signal domain $\boldsymbol{D}$, that is

$$\boldsymbol{s}=\sum_{i=1}^{N}\boldsymbol{d}_i h_i=\boldsymbol{D}\boldsymbol{h}, \tag{1}$$

where $\boldsymbol{h}=[h_1,h_2,\cdots,h_N]^T$ is the $K$-sparse coefficients vector ($K\ll N$), and $\boldsymbol{D}$ is an $N\times N$ orthogonal basis matrix with $\{\boldsymbol{d}_i, i=1,2,\cdots,N\}$ as its columns. Take a random measurement signal matrix $\boldsymbol{W}$ and then the received signal vector $\boldsymbol{y}=[y_1,\cdots,y_m,\cdots,y_M]^T$ can be written as

$$\begin{aligned}\boldsymbol{y}&=\boldsymbol{W}\boldsymbol{s}+\boldsymbol{z}\\&=\boldsymbol{W}\boldsymbol{D}\boldsymbol{h}+\boldsymbol{z}\\&=\boldsymbol{X}\boldsymbol{h}+\boldsymbol{z}\end{aligned} \tag{2}$$

where $\boldsymbol{X}=\boldsymbol{W}\boldsymbol{D}$ denotes a $M\times N$ sensing matrix as

$$\boldsymbol{X}=\begin{bmatrix}\boldsymbol{x}_1^T\\ \vdots\\ \boldsymbol{x}_m^T\\ \vdots\\ \boldsymbol{x}_M^T\end{bmatrix}=\begin{bmatrix}x_{11}&\cdots&x_{1n}&\cdots&x_{1N}\\ \vdots&\ddots&\vdots&\ddots&\vdots\\ x_{m1}&\cdots&x_{mn}&\cdots&x_{mN}\\ \vdots&\ddots&\ddots&\ddots&\vdots\\ x_{M1}&\cdots&x_{Mn}&\cdots&x_{MN}\end{bmatrix}, \tag{3}$$

and $\boldsymbol{z}=[z_1,\cdots,z_m,\cdots,z_M]^T$ is an additive white Gaussian noise (AWGN) with distribution $\mathcal{CN}(0,\sigma_n^2\boldsymbol{I}_M)$ and $\boldsymbol{I}_M$ denotes an $M\times M$ identity matrix. From the perspective of CS, the sensing matrix $\boldsymbol{X}$ satisfies the restricted isometry property (RIP) in overwhelming probability [9] so that the sparse signal $\boldsymbol{h}$ can be reconstructed correctly by NSS methods, e.g., BPDN [6] and OMP [7]. Take the BPDN as for the example to illustrate NSS realization approach. Since the sensing matrix $\boldsymbol{X}$ satisfies RIP of order $K$ with positive parameter $\delta_K\in(0,1)$, i.e., $\boldsymbol{X}\in\mathrm{RIP}(K,\delta_K)$ if

$$(1-\delta_K)\|\boldsymbol{h}\|_2^2\le\|\boldsymbol{X}\boldsymbol{h}\|_2^2\le(1+\delta_K)\|\boldsymbol{h}\|_2^2, \tag{4}$$

holds for all $\boldsymbol{h}$ having no more than $K$ nonzero coefficients. Then the unknown sparse vector $\boldsymbol{h}$ can be reconstructed by BPDN as

$$\tilde{\boldsymbol{h}}_{nss}=\arg\lim_{\tilde{h}}\left\{\frac{1}{2}\|\boldsymbol{y}-\boldsymbol{X}\boldsymbol{h}\|_2^2+\lambda\|\boldsymbol{h}\|_1\right\}, \tag{5}$$

where $\lambda$ denotes a regularization parameter which balances the mean-square error (MSE) term and sparsity of $\boldsymbol{h}$. If the mutual interference of sensing matrix $\boldsymbol{X}$ can be completely removed, then the theoretical Cramer-Rao lower bound (CRLB) of the NSS can be derived as [10]

$$\mathrm{CRLB}\{\tilde{\boldsymbol{h}}_{nss}\}=E\left\{\|\tilde{\boldsymbol{h}}_{nss}-\boldsymbol{h}\|_2\right\}=\frac{K\sigma_n^2}{N}. \tag{6}$$

## 3. Adaptive sparse sensing

We reconsider the above system model (2) with respect to adaptive sensing case. At observation side, $m$-th observed signal $y_m$ can be written as

$$y_m = \boldsymbol{h}^T \boldsymbol{x}_m + z_m, \tag{7}$$

for $m = 1, 2, \cdots, M$. The objective of ASS is to adaptively estimate the unknown sparse vector $\boldsymbol{h}$ using the sensing signal vector $\boldsymbol{x}_m$ and the observed signal $y_m$. Different from NSS approaches, we proposed an alternative ASS method using RZA-NLMF algorithm as shown in Fig. 2. Assume the $\tilde{y}_m(n) = \boldsymbol{x}_m^T \tilde{\boldsymbol{h}}(n)$ is an estimated observed signal which depends on signal estimator $\tilde{\boldsymbol{h}}(n)$ and hence the $n$-th observed signal error as $e_m(n) = y_m - \tilde{y}_m(n)$. Notice that the $e_m(n)$ is in correspondence with the $n$-th iterative error when using $m$-th sensing signal vector $\boldsymbol{x}_m$ and $m = \text{mod}(n, M)$. Notice that the $\text{mod}(\cdot)$ denotes a modulo function, for example, $\text{mod}(5,3) = 2$ and $\text{mod}(5,2) = 1$. First of all, the cost function of RZA-NLMF algorithm is constructed as

$$G(n) = \frac{1}{4} e_m^4(n) + \lambda_{ass} \sum_{i=1}^{N} \log\left(1 + \varepsilon |h_i|\right), \tag{8}$$

where $\lambda_{ass} > 0$ is a regularization parameter which trades off the sensing error and coefficients vector sparsity. $\varepsilon > 0$ denotes a reweighted factor which enhances to exploit the signal sparsity at each iteration. A figure example to show the relationship between reweighted factors and sparse constraint strength is given in Fig. 3. According to the cost function (8), the corresponding update equation can be derived as

$$\begin{aligned} \tilde{\boldsymbol{h}}(n+1) &= \tilde{\boldsymbol{h}}(n) - \mu_{iss} \frac{\partial G(n)}{\partial \tilde{\boldsymbol{h}}(n)} \\ &= \tilde{\boldsymbol{h}}(n) + \frac{\mu_{iss} e_m^3(n) \boldsymbol{x}_m}{\|\boldsymbol{x}_m\|_2^2 \left(\|\boldsymbol{x}_m\|_2^2 + e_m^2(n)\right)} - \frac{\rho \, \text{sgn}\left(\tilde{\boldsymbol{h}}(n)\right)}{1 + \varepsilon \left|\tilde{\boldsymbol{h}}(n)\right|} \\ &= \tilde{\boldsymbol{h}}(n) + \frac{\mu_{ass}(n) e_m(n) \boldsymbol{x}_m}{\|\boldsymbol{x}_m\|_2^2} - \frac{\rho \, \text{sgn}\left(\tilde{\boldsymbol{h}}(n)\right)}{1 + \varepsilon \left|\tilde{\boldsymbol{h}}(n)\right|}, \end{aligned} \tag{9}$$

where $\rho = \mu_{iss} \lambda / \varepsilon$ is a parameter which depends on initial step-size $\mu_{iss}$, regularization parameter $\lambda$ and threshold $\varepsilon$, respectively. In the second term of (9), if coefficient magnitudes of $\tilde{\boldsymbol{h}}(n)$ are smaller than $1/\varepsilon$, then these small coefficients will be replaced by zeros in high probability [11]. Here, it is worth noting that $\mu_{ass}(n)$ is a variable step-size

$$\mu_{ass}(n) = \frac{\mu_{iss} e_m^2(n)}{\|\boldsymbol{x}_m\|_2^2 + e_m^2(n)}, \tag{10}$$

which depends on three factors: initial step-size $\mu_{iss}$, input signal $\boldsymbol{x}_m$ and update iterative error $e_m(n)$. Since $\mu_{iss}$ is given initial steps-size and $\boldsymbol{x}_m$ is random scaling input signal, hence, $\mu_{ass}$ in Eq. (10) can also be rewritten as

$$\mu_{ass}(n)=\frac{\mu_{iss}}{\left\|\boldsymbol{x}_m\right\|_2^2 / e_m^2(n)+1}, \tag{11}$$

which is a variable step-size (VSS) which is adaptive change as square sensing error $e_m^2(n)$, smaller error incurs the smaller step-size to ensure the stability of the gradient descend while larger error yields larger step-size to accelerate the convergence speed of this algorithm [12]. According to the update equation in (9), our proposed ASS method can be concluded in **Algorithm 1**.

As for the trademark of the performance comparisons, CRLB of the proposed ASS method is derived in the subsequent. The signal error is defined as $\boldsymbol{v}(n):=\tilde{\boldsymbol{h}}(n)-\boldsymbol{h}$ and $e(n)$ can be written as $e_m(n)=z_m-\boldsymbol{v}^T(n)\boldsymbol{x}_m$. To simply derive the CRLB, four assumptions are considered in the subsequent analysis: 1) the input signal $\boldsymbol{x}_m$ and noise $z_m$ are mutually independent; 2) each row $\boldsymbol{x}_m$ of the signal matrix $\boldsymbol{X}$ is random independent with zero mean and random Gaussian variance $\sigma^2\boldsymbol{I}_N$; 3) noise $z_m$ is random independent with zero mean and variance $\sigma_n^2$; 4) $\tilde{\boldsymbol{h}}(n)$ is independent of $\boldsymbol{X}$. Assume that the $n$-th adaptive receive error $e(n)$ sufficient small so that $e_m^2(n)\ll\boldsymbol{x}_m$, hence $\mu_{ass}=\mu_{iss}\,e_m^2(n)/\boldsymbol{x}_m$, according to (9), the $n$-th update signal error $\boldsymbol{v}(n+1)$ can be written as

$$\boldsymbol{v}(n+1)=\boldsymbol{v}(n)+\frac{\mu_{iss}e_m^3(n)\boldsymbol{x}_m}{\left\|\boldsymbol{x}_m\right\|_2^2}-\frac{\rho\,\mathrm{sgn}\left(\tilde{\boldsymbol{h}}(n)\right)}{1+\varepsilon\left|\tilde{\boldsymbol{h}}(n)\right|}, \tag{12}$$

where $e_m^3(n)$ can be expended as

$$\begin{aligned} e_m^3(n)&=\left(z_m-\boldsymbol{v}^T(n)\boldsymbol{x}_m\right)^3 \\ &=z_m^3-3z_m^2\boldsymbol{v}^T(n)\boldsymbol{x}_m+3z_m\left(\boldsymbol{v}^T(n)\boldsymbol{x}_m\right)^2-\left(\boldsymbol{v}^T(n)\boldsymbol{x}_m\right)^3. \end{aligned} \tag{13}$$

Substituting (13) into (12), $\boldsymbol{v}(n+1)$ can be further represented as

$$\begin{aligned} \boldsymbol{v}(n+1)=\boldsymbol{v}(n)&+\frac{\mu_{iss}z_m^3\boldsymbol{x}_m}{\left\|\boldsymbol{x}_m\right\|_2^2}-\frac{3\mu_{iss}z_m^2\left(\boldsymbol{v}^T(n)\boldsymbol{x}_m\right)\boldsymbol{x}_m}{\left\|\boldsymbol{x}_m\right\|_2^2} \\ &+\frac{3\mu_{iss}z_m\left(\boldsymbol{v}^T(n)\boldsymbol{x}_m\right)^2\boldsymbol{x}_m}{\left\|\boldsymbol{x}_m\right\|_2^2}-\frac{\mu_{iss}\left(\boldsymbol{v}^T(n)\boldsymbol{x}_m\right)^3\boldsymbol{x}_m}{\left\|\boldsymbol{x}_m\right\|_2^2}-\frac{\rho\,\mathrm{sgn}\left(\tilde{\boldsymbol{h}}(n)\right)}{1+\varepsilon\left|\tilde{\boldsymbol{h}}(n)\right|}. \end{aligned} \tag{14}$$

Hence, the steady-state mean square error (MSE) can be derived as

$$
\begin{aligned}
E\left[\boldsymbol{v}^T(n+1)\boldsymbol{v}(n+1)\right] = {} & E\left[\boldsymbol{v}^T(n)\boldsymbol{v}(n)\right] + \mu_{nss}^2 E\left[z_m^6 \big/ \|\boldsymbol{x}_m\|_2^2\right] \\
& + 9\mu_{iss}^2 E\left[\frac{z_m^4\left(\boldsymbol{v}^T(n)\boldsymbol{x}_m\right)^2}{\|\boldsymbol{x}_m\|_2^2}\right] + 9\mu_{nss}^2 E\left[\frac{z_m^2\left(\boldsymbol{v}^T(n)\boldsymbol{x}_m\right)^4}{\|\boldsymbol{x}_m\|_2^2}\right] \\
& + \mu_{iss}^2 E\left[\frac{z_m^2\left(\boldsymbol{v}^T(n)\boldsymbol{x}_m\right)^6}{\|\boldsymbol{x}_m\|_2^2}\right] + \rho^2 E\left[\frac{\operatorname{sgn}\left(\tilde{\boldsymbol{h}}^T(n)\right)\operatorname{sgn}\left(\tilde{\boldsymbol{h}}(n)\right)}{\left(1+\varepsilon\left|\tilde{\boldsymbol{h}}(n)\right|\right)^2}\right] \\
& + 2\mu_{iss} E\left[\frac{z_m^3\boldsymbol{v}^T(n)\boldsymbol{x}_m}{\|\boldsymbol{x}_m\|_2^2}\right] - 6\mu_{iss} E\left[\frac{z_m^2\left(\boldsymbol{v}^T(n)\boldsymbol{x}_m\right)^2}{\|\boldsymbol{x}_m\|_2^2}\right] \\
& + 6\mu_{iss} E\left[\frac{z_m\left(\boldsymbol{v}^T(n)\boldsymbol{x}_m\right)^3}{\|\boldsymbol{x}_m\|_2^2}\right] - 2\mu_{iss} E\left[\frac{\left(\boldsymbol{v}^T(n)\boldsymbol{x}_m\right)^4}{\|\boldsymbol{x}_m\|_2^2}\right] \\
& - 2\rho E\left[\frac{\boldsymbol{v}^T(n)\operatorname{sgn}\left(\tilde{\boldsymbol{h}}(n)\right)}{1+\varepsilon\left|\tilde{\boldsymbol{h}}(n)\right|}\right] - 6\mu_{iss}^2 E\left[\frac{z_m^5\boldsymbol{v}^T(n)\boldsymbol{x}_m}{\|\boldsymbol{x}_m\|_2^2}\right] \\
& + 6\mu_{iss}^2 E\left[\frac{z_m^4\left(\boldsymbol{v}^T(n)\boldsymbol{x}_m\right)^2}{\|\boldsymbol{x}_m\|_2^2}\right] - 2\mu_{iss}^2 E\left[\frac{z_m^3\left(\boldsymbol{v}^T(n)\boldsymbol{x}_m\right)^3}{\|\boldsymbol{x}_m\|_2^2}\right] \\
& - 2\rho\mu_{iss} E\left[\frac{z_m^3\boldsymbol{x}_m^T\operatorname{sgn}\left(\tilde{\boldsymbol{h}}(n)\right)}{\|\boldsymbol{x}_m\|_2^2\left(1+\varepsilon\left|\tilde{\boldsymbol{h}}(n)\right|\right)}\right] - 18\mu_{iss}^2 E\left[\frac{z_m^3\left(\boldsymbol{v}^T(n)\boldsymbol{x}_m\right)^3}{\|\boldsymbol{x}_m\|_2^2}\right] \\
& + 6\mu_{iss}^2 E\left[\frac{\left(\boldsymbol{v}^T(n)\boldsymbol{x}_m\right)^4}{\|\boldsymbol{x}_m\|_2^2}\right] + 6\rho\mu_{iss} E\left[\frac{z_m^2\left(\boldsymbol{v}^T(n)\boldsymbol{x}_m\right)\boldsymbol{x}_m^T\operatorname{sgn}\left(\tilde{\boldsymbol{h}}(n)\right)}{\|\boldsymbol{x}_m\|_2^2\left(1+\varepsilon\left|\tilde{\boldsymbol{h}}(n)\right|\right)}\right].
\end{aligned} \tag{15}
$$

Based on above mentioned independent assumptions and ideal Gaussian noise assumption [13], we can get the following approximations

$$E[z_m] = E[z_m^3] = E[z_m^5] = 0, \tag{16}$$

$$E[z_m^4] = 3\sigma_n^4, \tag{17}$$

$$E[z_m^6] = 15\sigma_n^6, \tag{18}$$

$$E[\boldsymbol{x}_m^T\boldsymbol{x}_m] = N\sigma^2. \tag{19}$$

Due to the independence between $\boldsymbol{x}_m$ and $\boldsymbol{v}(n)$, $\{\boldsymbol{v}^T(n)\boldsymbol{x}_m\}$ satisfies zero-mean Gaussian distribution, that is $E[\boldsymbol{v}^T(n)\boldsymbol{x}_m] = 0$ [13]. Hence, we can also get following approximations

$$E\left[\left(\boldsymbol{v}^T(n)\boldsymbol{x}_m\right)^2 |\boldsymbol{v}(n)\right] = \sigma^2 E\left[\boldsymbol{v}^T(n)\boldsymbol{v}(n)\right], \tag{20}$$

$$E\left[\left(\boldsymbol{v}^T(n)\boldsymbol{x}_m\right)^4 |\boldsymbol{v}(n)\right] = 3\sigma^4 E\left[\boldsymbol{v}^T(n)\boldsymbol{v}(n)\right]^2, \tag{21}$$

$$E\left[\left(\boldsymbol{v}^T(n)\boldsymbol{x}_m\right)^6 |\boldsymbol{v}(n)\right] = 15\sigma^6 E\left[\boldsymbol{v}^T(n)\boldsymbol{v}(n)\right]^3. \tag{22}$$

By neglecting the random fluctuations in $\boldsymbol{v}^T(n)\boldsymbol{v}(n)$ and using approximation equation $\boldsymbol{v}^T(n)\boldsymbol{v}(n) \approx E[\boldsymbol{v}^T(n)\boldsymbol{v}(n)] = b(n)$, substitute (16)-(22) into (15) which can be simplified as

$$
\begin{aligned}
b(n+1) &= \left[1 + \frac{27\mu_{iss}^2\sigma_n^4 - 6\mu_{iss}\sigma_n^2}{N}\right] b(n) \\
&+ \left[\frac{27\mu_{iss}^2\sigma_n^2\sigma^2 - 2\mu_{iss}3\sigma^2}{N} + \frac{18\mu_{iss}^2\sigma^2}{N\sigma^2}\right] b^2(n) \\
&+ \frac{15\mu_{iss}^2\sigma_n^2\sigma^4}{N} b^3(n) - \frac{15\mu_{iss}^2\sigma_n^6}{N\sigma^2} + \phi(n),
\end{aligned}
\tag{23}
$$

where $\phi(n)$ is incurred by the last term of (12) and it is expressed by

$$
\begin{aligned}
\phi(n) &= \frac{6\rho\mu_{iss}\sigma_n^2}{N} E\left[\frac{\left(\boldsymbol{v}^T(n)\boldsymbol{x}_m\right)\boldsymbol{x}_m^T \operatorname{sgn}\left(\tilde{\boldsymbol{h}}(n)\right)}{1 + \varepsilon\left|\tilde{\boldsymbol{h}}(n)\right|}\right] \\
&+ \rho^2 E\left[\frac{\operatorname{sgn}\left(\tilde{\boldsymbol{h}}^T(n)\right)\operatorname{sgn}\left(\tilde{\boldsymbol{h}}(n)\right)}{\left(1 + \varepsilon\left|\tilde{\boldsymbol{h}}(n)\right|\right)^2}\right] - 2\rho E\left[\frac{\boldsymbol{v}^T(n)\operatorname{sgn}\left(\tilde{\boldsymbol{h}}(n)\right)}{1 + \varepsilon\left|\tilde{\boldsymbol{h}}(n)\right|}\right].
\end{aligned}
\tag{24}
$$

Since the adaptive update square error $b(n)$ is too small (i.e., $b(n) \ll 1$), hence, higher than two-order errors are considered zero, i.e., $b^2(n) = 0$ and $b^3(n) = 0$. The MSE can be derived from (23) as

$$
b(\infty) = \frac{5\mu_{iss}\sigma_n^4}{9\mu_{iss}\sigma_n^2\sigma^2 - 2\sigma^2} - \frac{N\phi(\infty)}{27\mu_{iss}^2\sigma_n^4 - 6\mu_{iss}\sigma_n^2}. \tag{25}
$$

Assume that ideal reconstruction vector $\tilde{\boldsymbol{h}}(n)$ can be obtained, then one can get $\lim_{n\to\infty}\left\|\tilde{\boldsymbol{h}}(n)\right\|_1 = \|\boldsymbol{h}\|_1$ and $\lim_{n\to\infty} \operatorname{sgn}(\tilde{\boldsymbol{h}}^T(n))\operatorname{sgn}(\tilde{\boldsymbol{h}}(n)) = K$, where $K$ denotes the number of nonzero coefficients in $\boldsymbol{h}$. Hence, $\phi(\infty)$ in (25) can be derived as

$$
\begin{aligned}
\phi(\infty) &= \lim_{n\to\infty}\phi(n) \\
&= \lim_{n\to\infty}\left\{\frac{6\rho\mu_{iss}\sigma_n^2\sigma^2}{N} - 2\rho\right\} E\left[\frac{\left(\tilde{\boldsymbol{h}}(n) - \boldsymbol{h}\right)^T \operatorname{sgn}\left(\tilde{\boldsymbol{h}}(n)\right)}{1 + \varepsilon\left|\tilde{\boldsymbol{h}}(n)\right|}\right] + \rho^2 E\left[\frac{\operatorname{sgn}\left(\tilde{\boldsymbol{h}}^T(n)\right)\operatorname{sgn}\left(\tilde{\boldsymbol{h}}(n)\right)}{\left(1 + \varepsilon\left|\tilde{\boldsymbol{h}}(n)\right|\right)^2}\right] \\
&= \lim_{n\to\infty}\left\{\frac{6\rho\mu_{iss}\sigma_n^2\sigma^2}{N} - 2\rho\right\} E\left[\left\|\frac{\tilde{\boldsymbol{h}}(n)}{1 + \varepsilon\left|\tilde{\boldsymbol{h}}(n)\right|}\right\|_1 - \left\|\frac{\boldsymbol{h}}{1 + \varepsilon\left|\tilde{\boldsymbol{h}}(n)\right|}\right\|_1\right] + \rho^2 E\left[\frac{\operatorname{sgn}\left(\tilde{\boldsymbol{h}}^T(n)\right)\operatorname{sgn}\left(\tilde{\boldsymbol{h}}(n)\right)}{\left(1 + \varepsilon\left|\tilde{\boldsymbol{h}}(n)\right|\right)^2}\right] \\
&\le \rho^2 K
\end{aligned}
\tag{26}
$$

Finally, the CRLB of the proposed ASS can be obtained as

$$
\mathrm{CRLB}\{\tilde{\boldsymbol{h}}_{ass}\} = b(\infty) = \frac{5\mu_{iss}\sigma_n^4}{9\mu_{iss}\sigma_n^2\sigma^2 - 2\sigma^2} - \frac{\rho^2 NK}{27\mu_{iss}^2\sigma_n^4 - 6\mu_{iss}\sigma_n^2}. \tag{27}
$$

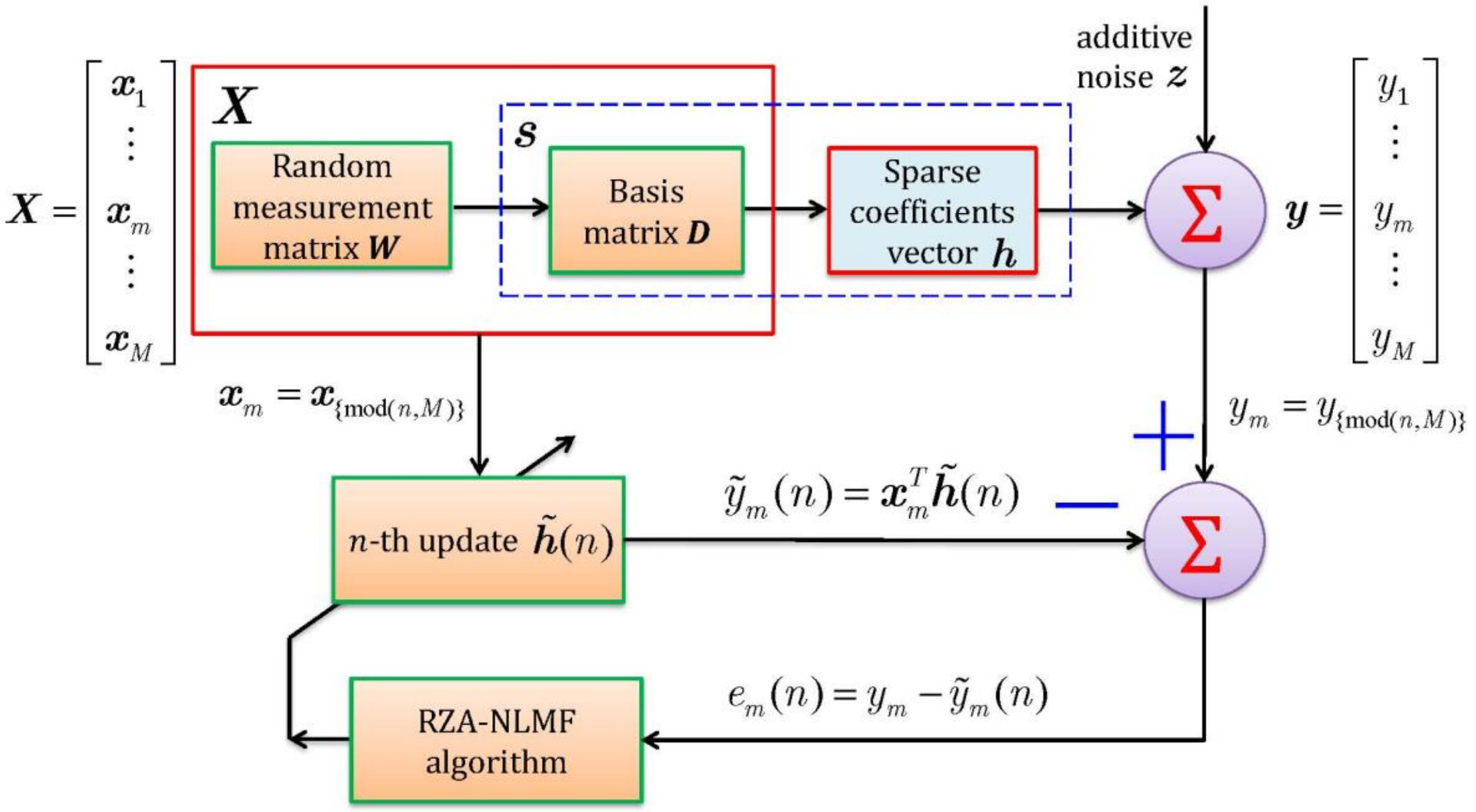

**Figure 2. RZA-NLMF algorithm for ASS.**

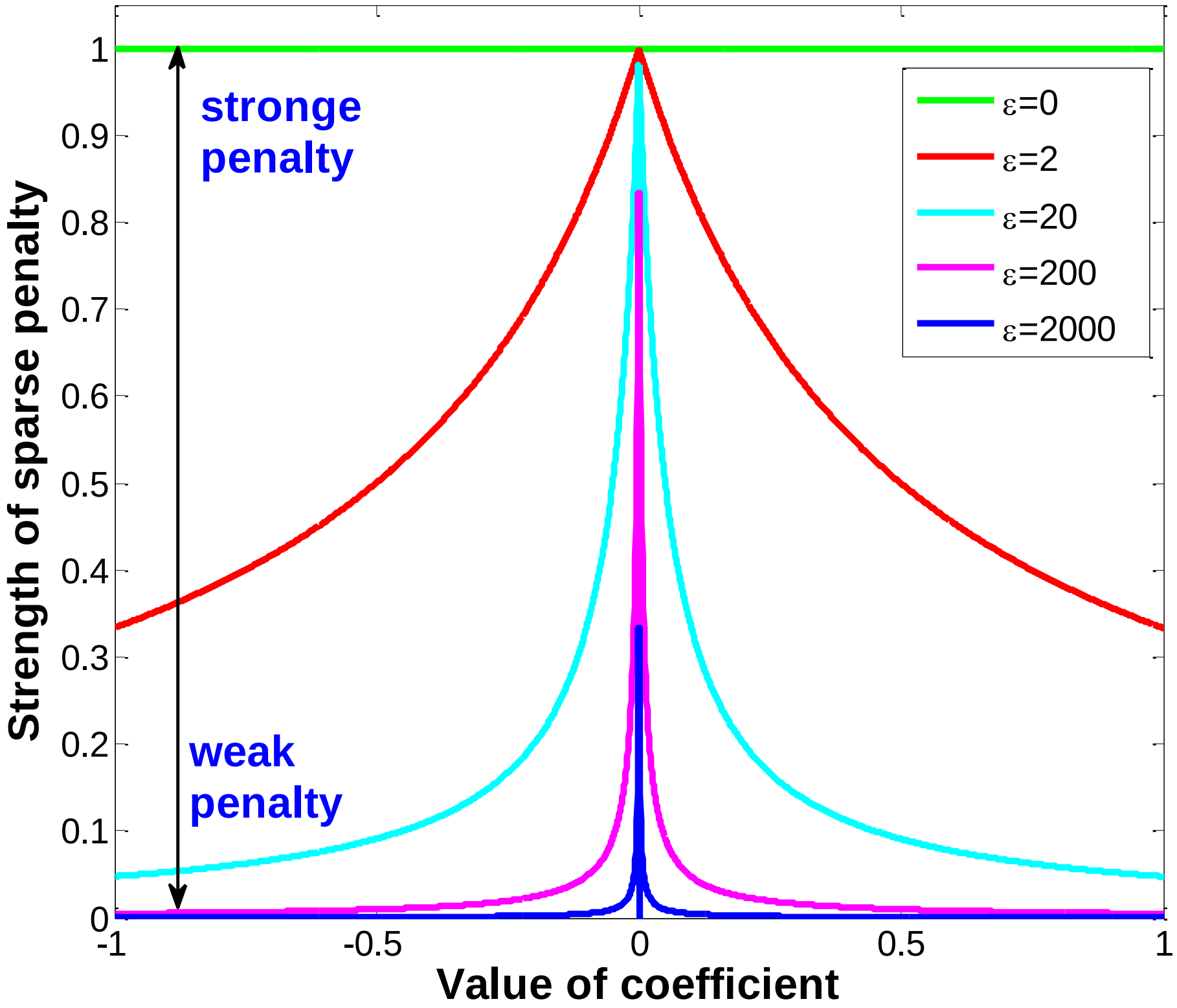

**Figure 3. Sparse constraint strength comparison using different reweights.**

**Algorithm 1. ASS using RZA-NLMF algorithm for solving CS problems.**

**Input**: Random sensing matrix $\boldsymbol{X}$, observation signal vector $\boldsymbol{y}$.

**Output**: $\tilde{\boldsymbol{h}}$.

(1) Initialize $\tilde{\boldsymbol{h}}(0)=\boldsymbol{0}$, $n=1$, set step-size $\mu_{iss}$, reweighted factor $\varepsilon$, regularization parameter $\lambda$.

(2) Send data $\boldsymbol{x}_m$ and $y_m$ to RZA-NLMD filter, where $m=\text{mod}(n,M)+1$.

(3) While stop condition $\|\tilde{\boldsymbol{h}}(n+1)-\tilde{\boldsymbol{h}}(n)\|_2<\zeta$ or $n>n_{\max}$ where $\zeta>0$ is a given error tolerance and $n_{\max}$ is a given maximum iteration number.

(4) Determine the input signal $\boldsymbol{x}_m$ and observation signal $y_m$

(5) Calculate error $e_m(n)$ as $e_m(n)=y_m-\boldsymbol{x}_m^T\tilde{\boldsymbol{h}}(n)$.

(6) Update $\tilde{\boldsymbol{h}}(n+1)=\tilde{\boldsymbol{h}}(n)+\mu_{ass}(n)e_m(n)\boldsymbol{x}_m\Big/\|\boldsymbol{x}_m\|_2^2+\rho\,\text{sgn}\left(\tilde{\boldsymbol{h}}(n)\right)\Big/\left(1+\varepsilon\left|\tilde{\boldsymbol{h}}(n)\right|\right)$

(7) Iteration number increases by one $n=n+1$

(8) End while

## 4. Computer Simulations

In this section, the proposed ASS approach using RZA-NLMF algorithm is evaluated. For achieving average performance, 1000 independent Monte-Carlo runs are adopted. For easy evaluating the effectiveness of the proposed approach, signal representation domain $\boldsymbol{D}$ is assumed as an identity matrix $\boldsymbol{I}_{N\times N}$ and unknown signal $\boldsymbol{s}$ is set as sparse directly. Sensing matrix is equivalent to random measurement matrix, i.e., $\boldsymbol{X}=\boldsymbol{W}$. For ensuring $\boldsymbol{X}$ satisfies the RIP, $\boldsymbol{W}$ is set as random Gaussian matrix [9]. Then, sparse coefficient vector $\boldsymbol{h}$ equals to $\boldsymbol{s}$. The detail simulation parameters are listed in Tab. 1. Notice that each nonzero coefficient of $\boldsymbol{h}$ follows random Gaussian distribution as $\mathcal{CN}(0,\sigma^2)$ and their positions are randomly allocated within the signal length of $\boldsymbol{h}$ which is subject to $E\{\|\boldsymbol{h}\|_2^2\}=1$, where $E\{\cdot\}$ denotes the expectation operator. The output signal-to-noise ratio (SNR) is defined as 20log $(E_s/\sigma_n^2)$, where $E_s=1$ is the unit transmission power. All of the step sizes and regularization parameters are listed in Tab. I. The estimation performance is evaluated by average mean square error (MSE) which is defined by

$$\text{Average MSE}\left\{\tilde{\boldsymbol{h}}(n)\right\}:=E\left\{\left\|\boldsymbol{h}-\tilde{\boldsymbol{h}}(n)\right\|_2^2\right\}, \tag{28}$$

where $\boldsymbol{h}$ and $\tilde{\boldsymbol{h}}(n)$ are the actual channel vector and its $n$-th iterative adaptive channel estimator, respectively. According to our previous work [8], regularization parameter for RZA-NLMF is set as $\lambda=5\times10^{-8}$ so that it can exploit signal sparsity robustly. Since the RZA-NLMF-based ASS method depends highly on the reweighted factor $\varepsilon$, hence, we first select the reasonable factor $\varepsilon$ by virtual of Monte Carlo. Later, we compare the proposed method with two typical NSS ones, i.e., BPDN [6] and OMP [7].

## 4.1. Reweighted factor selection

Since the RZA-LMSF algorithm depends highly on reweighted factor. Hence, selection of the robust reweighted factor for different noise environments and different signal sparsities is typical important for the RZA-LMSF algorithm. By means of Monte Carlo method, performance curves of the proposed ASS method with different reweighted factors $\varepsilon \in \{2,20,200,2000,20000\}$ with respect to different number of nonzero coefficients $K \in \{2,6,10\}$ and different SNR regimes (5dB and 10dB) are depicted in Figs. 4~7. Under the simulation setup considered, RZA-NLMF using $\varepsilon = 2000$ can achieve robust performance in different cases as shown in Figs. 4~7. From the four figures, one can find that sparser signal requires larger reweighted factor but no more than 20000 in this system. This is concise with the fact that stronger sparse penalty not only exploits more sparse information but also mitigates more noise interference.

**Tab. 1. Simulation parameters.**

| **Parameters** | **Values** |
|---|---|
| Signal length | $N = 40$ |
| Measurement length | $M = 20$ |
| Sensing matrix | Random Gaussian distribution |
| No. of nonzero coefficients | $K \in \{2,6,10\}$ |
| Distribution of nonzero coefficients | Random Gaussian |
| Signal-to-noise ratio (SNR) | (0dB,12dB) |
| Initial step-size: $\mu_{iss}$ | 1.5 |
| Regularization parameter: $\lambda$ | $5 \times 10^{-8}$ |
| Re-weighted factor: $\varepsilon$ | 2000 |

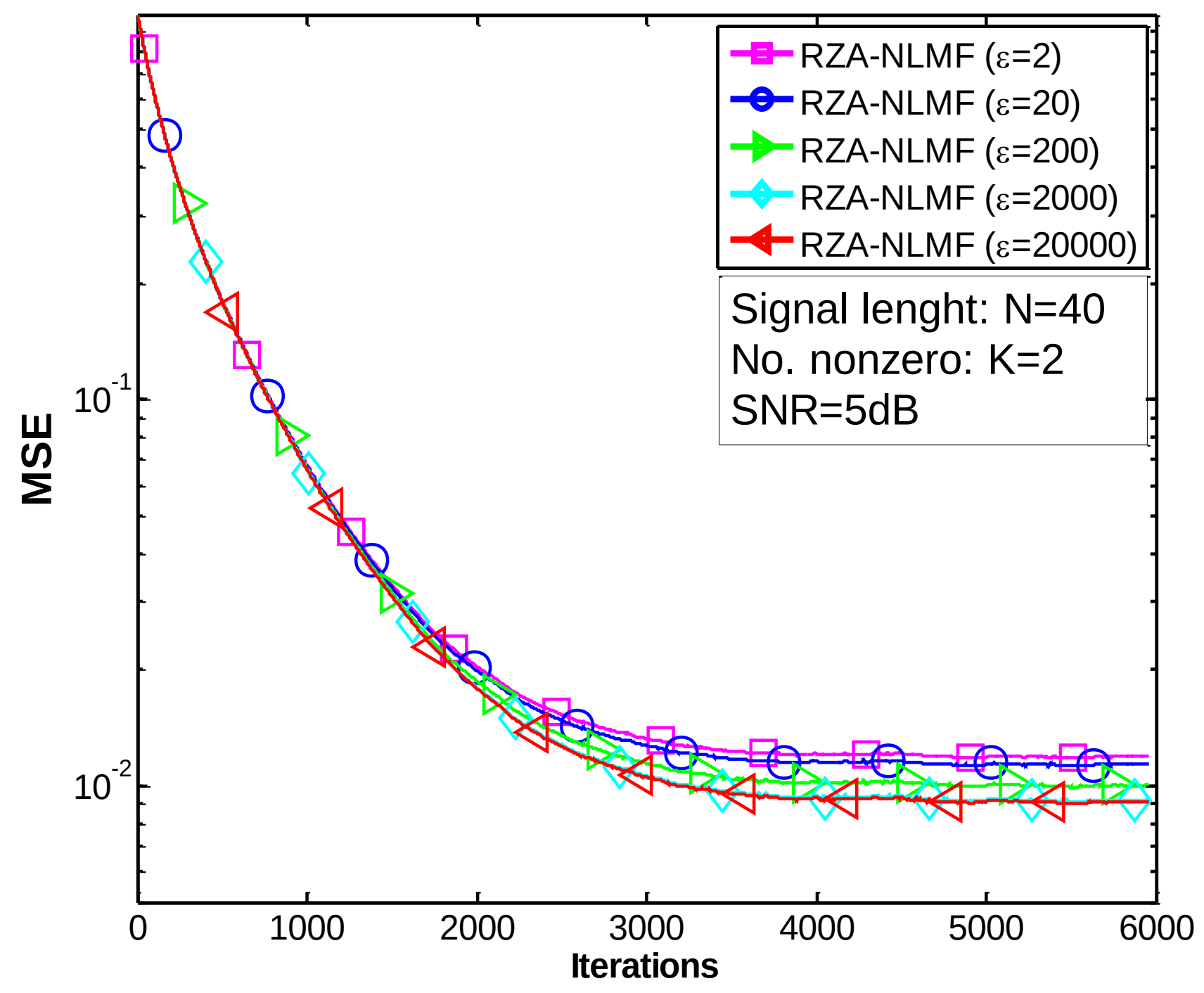

**Figure 4. RZA-NLMF performance verses reweighted factors (K=2 and SNR=5dB).**

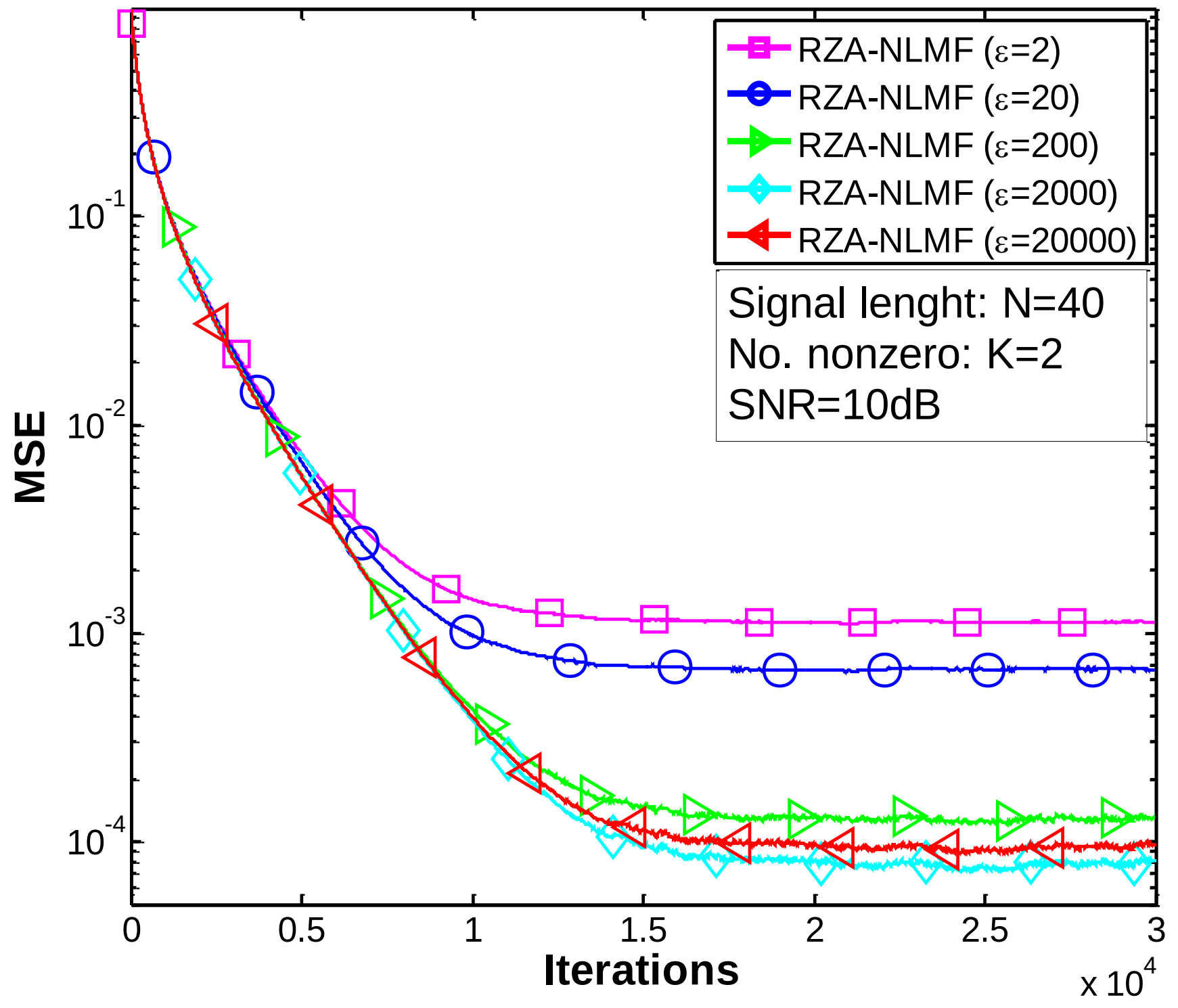

**Figure 5. RZA-NLMF performance verses reweighted factors (K=2 and SNR=10dB).**

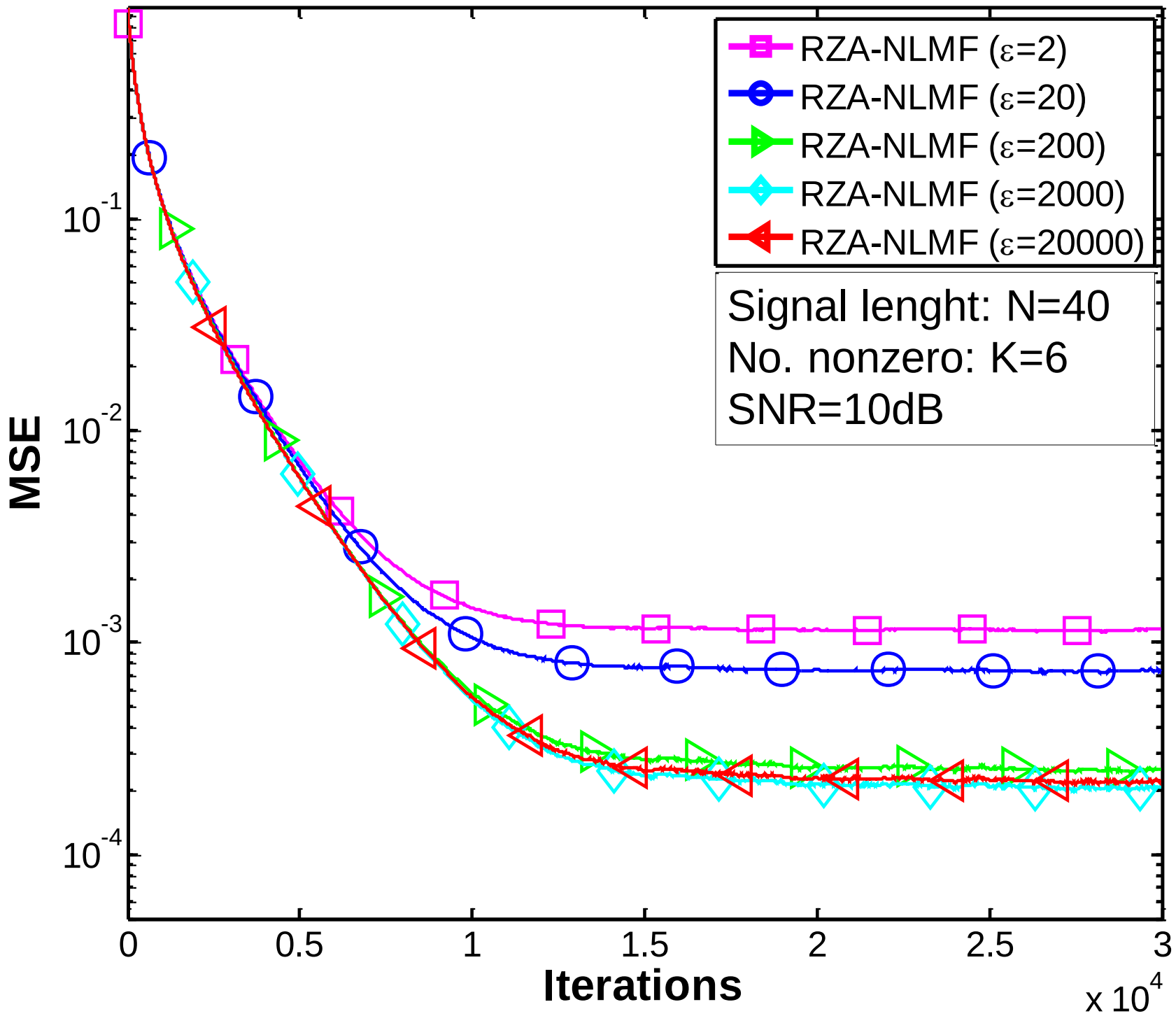

**Figure 6. RZA-NLMF performance verses reweighted factors (K=6 and SNR=10dB).**

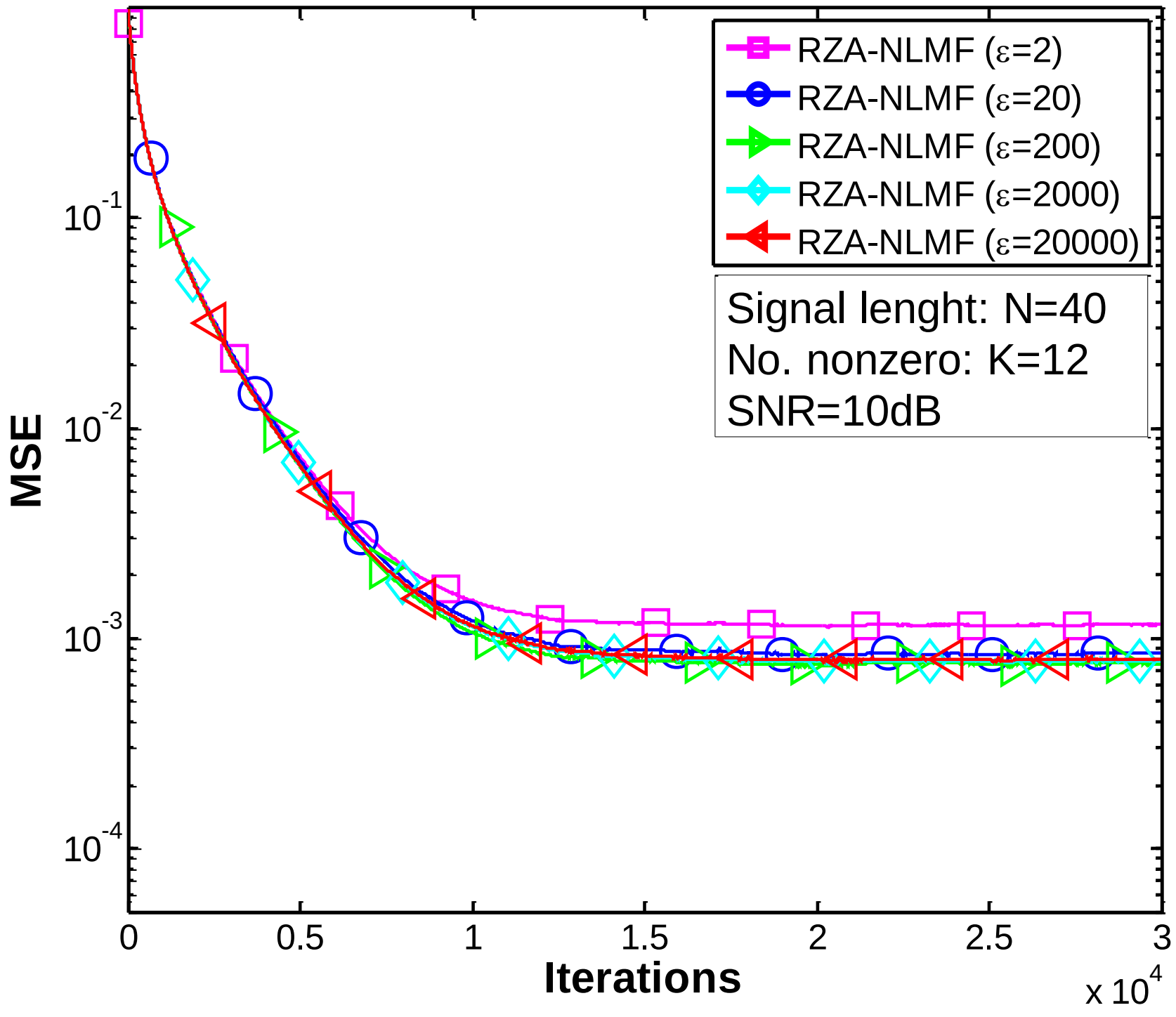

**Figure 7. RZA-NLMF performance verses reweighted factors (K=10 and SNR=10dB).**

## 4.2. Performance comparisons with NSS

Two experiments of ASS are verified in performance comparisons with conventional NSS methods (e.g., BPDN [6] and OMP [7]). In the first experiment, ASS method is evaluated in the case of $\mathrm{SNR} = 10\,\mathrm{dB}$ as shown in Fig. 8. On the one hand, according to this figure, we can find that the proposed ASS method using RZA-NLMF algorithm achieves much lower MSE performance than NSS methods and even if its CRLB. The existing big performance gap between ASS and NSS is that ASS using RZA-NLMF not only exploits the signal sparsity but also mitigates the noise interference using high-order error statistis for adaptive error updating. On the other hand, we can also find that ASS depends on the signal sparseness. That is to say, for sparser signal, ASS can exploit more signal structure information as for prior information and vice versa. In the second experiment, number of nonzero coefficients is fixed as $K = 2$ as shown in Fig. 9. It is easy to find that our proposed ASS is much better than conventional NSS as the SNR increasing.

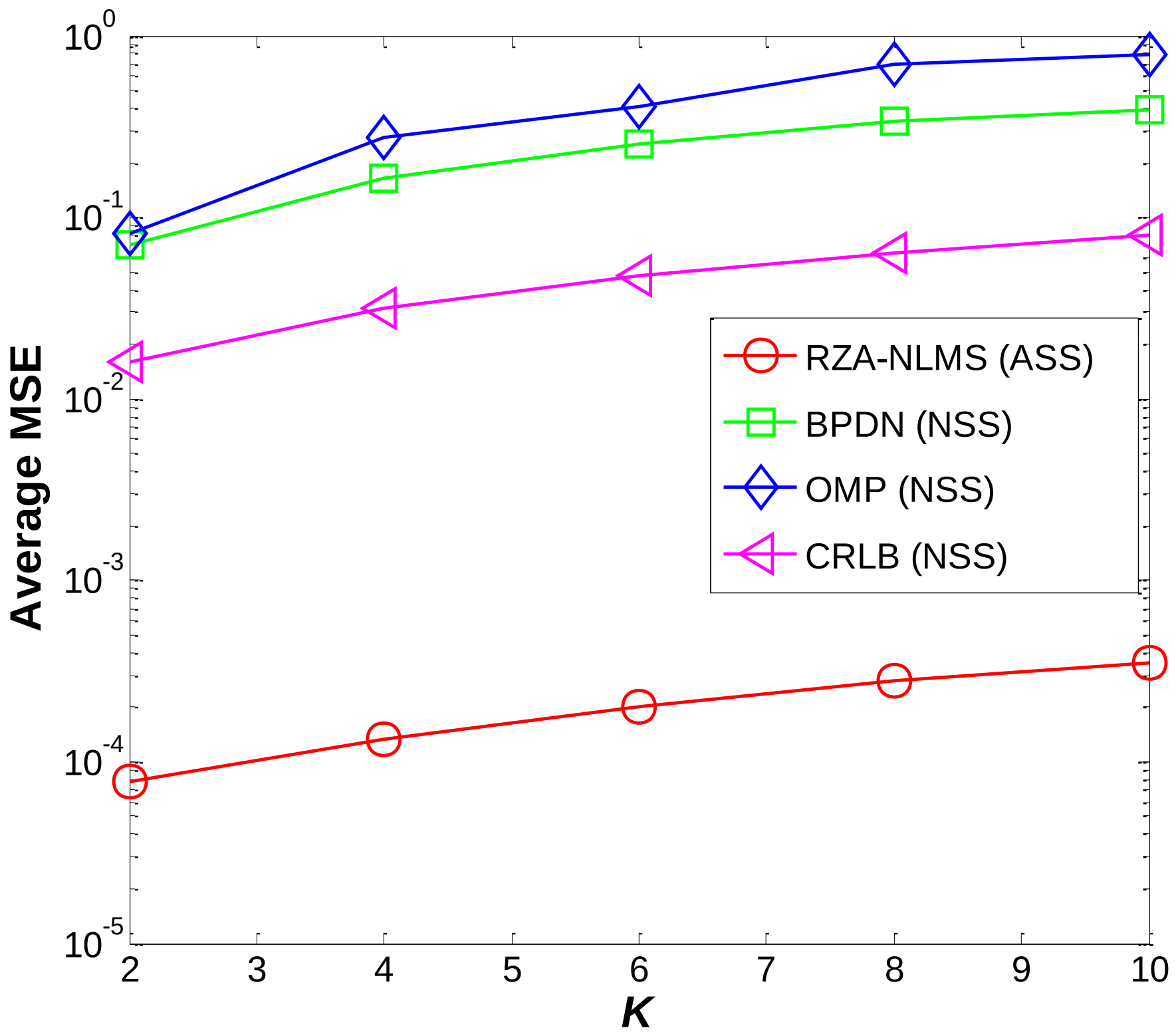

**Figure 8. Performance comparisons verses signal sparisty.**

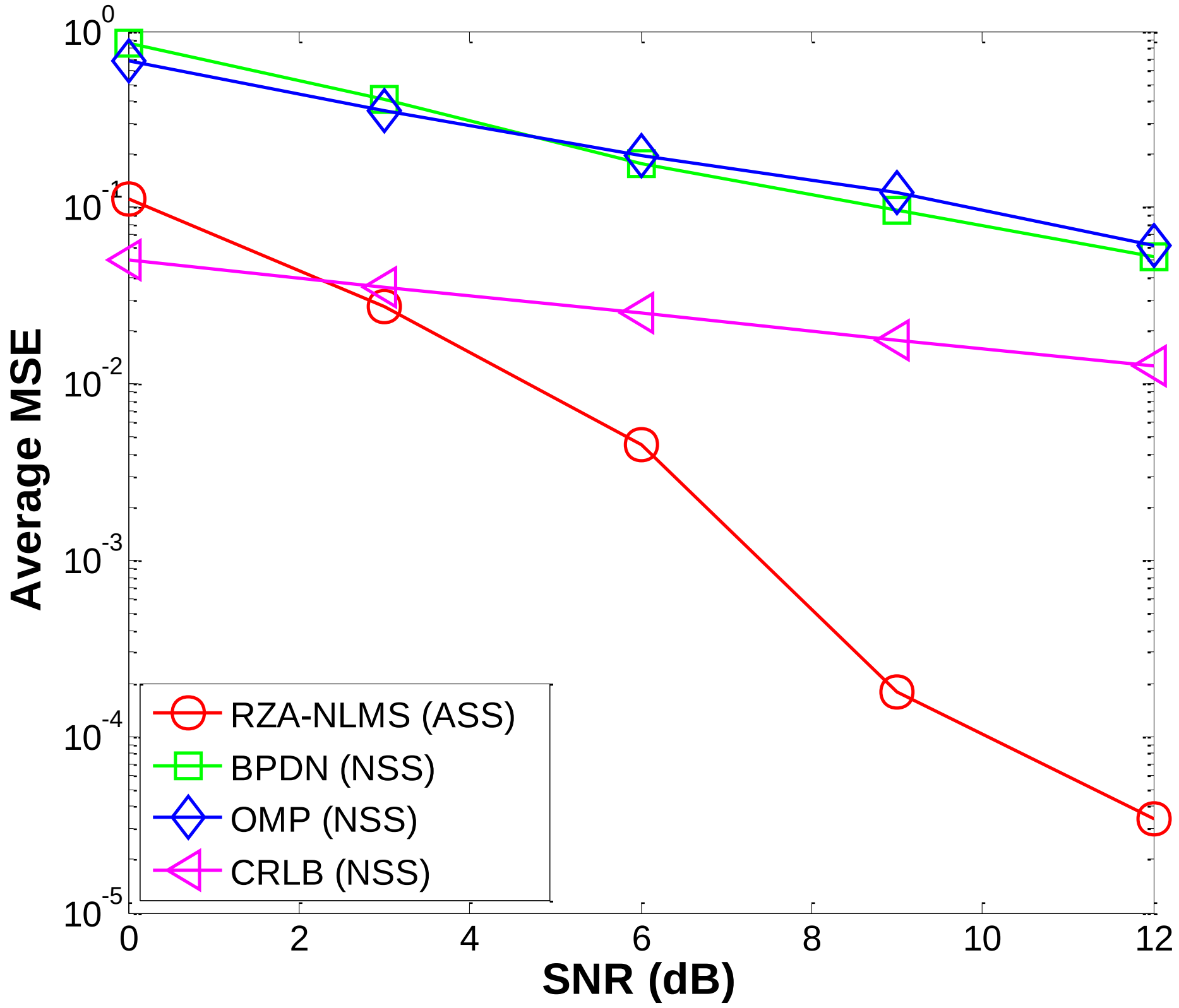

**Figure 9. Performance comparisons verses SNR.**

## 5. Conclusion

In this paper, we proposed an ASS method using RZA-NLMF algorithm for dealing with the CS problems. First, we decided the reweighted factor and regularization parameter for the proposed algorithm by virtual of Monte Carlo method. Later, based on update equation of the RZA-NLMF, CRLB of ASS was also derived based on the random independent assumptions. Finally, several representative simulations have been given to show that proposed method achieves much better MSE performance than NSS with respect to different signal sparsity, especially in the case of low SNR regime.

## 6. Acknowledgments

This work was supported by a grant-in-aid from Japan Society for the Promotion of Science (JSPS) (grant number 24·02366).